\documentclass[useAMS,usenatbib]{mnras}
\usepackage{natbib}
\usepackage{graphicx}
\graphicspath{ {\images} }
\usepackage{savesym}
\usepackage{listings}
\usepackage{multirow}
\usepackage[intlimits]{amsmath}
\savesymbol{iint}
\usepackage{txfonts}
\usepackage{bm}
\usepackage{amssymb}
\usepackage{amsmath}
\usepackage{tabularx}
\usepackage{color}
\usepackage{comment}
\usepackage{float}

\title[Twisted magnetar magnetospheres]{Twisted magnetar magnetospheres}
\author[D. Ntotsikas, K.N. Gourgouliatos, I. Contopoulos, S.K. Lander]{{D. Ntotsikas$^{1}$\thanks{Email: d.ntotsikas@upnet.com}}, K. N. Gourgouliatos$^{1}$\thanks{Email: kngourg@upatras.gr}, I. Contopoulos$^{2}$\thanks{Email: icontop@academyofathens.gr} and S. K. Lander${}^3$ \vspace{0.4cm}\\
\parbox{\textwidth}{$^{1}$ Department of Physics, University of Patras, Patras, Rio, 26504, Greece.\\
${}^2$ Research Center for Astronomy and Applied Mathematics, Academy of Athens, Athens, 11527, Greece.\\
${}^3$ Physics, Faculty of Science, University of East Anglia, Norwich, NR4 7TJ, U.K.
} } 
\begin{document}

\date{Accepted -. Received -; in original form -}
\pagerange{\pageref{firstpage}--\pageref{lastpage}} \pubyear{-}

\maketitle

\label{firstpage}

\begin{abstract} 
    Magnetar magnetospheres are strongly twisted, and are able to power sudden energetic events through the rapid release of stored electromagnetic energy. In this paper, we investigate twisted relativistic force-free axisymmetric magnetospheres of rotating neutron stars. We obtain numerical solutions of such configurations using the method of simultaneous relaxation for the magnetic field inside and outside the light-cylinder. 
    We introduce a toroidal magnetic field in the region of closed field-lines that is associated with a poloidal electric current distribution in that region, and explore various mathematical expressions for that distribution. We find that, by increasing the twist, a larger fraction of magnetic field-lines crosses the light-cylinder and opens up to infinity, thus increasing the size of the polar caps and enhancing the spin-down rate. We also find that, for moderately to strongly twisted magnetospheres, the region of closed field-lines ends at some distance inside the light-cylinder. We discuss the implications of these solutions on the variation of magnetar spin-down rates, moding and nulling of pulsars, the relation between the angular shear and the twist 
    and the overall shape of the magnetosphere.
\end{abstract}



\section{Introduction}

A major part of pulsar study is the modeling of their magnetosphere, that is the neutron star atmosphere with its non-trivial distribution of magnetic field-lines, charge density, and electric currents.
In a broader astrophysical context, early studies of magnetospheres focused initially on vacuum fields \citep{deutsch1955electromagnetic}. The work of
\cite{Goldreich:1969} established the framework for a force-free magnetosphere of a relativistically rotating pulsar, initiating the effort for analytical and numerical solutions. The first solution of an aligned, axisymmetric and rotating neutron star with a smooth transition of the magnetic field-lines from the inner magnetosphere to the outer magnetosphere through the light-cylinder  \citep{contopoulos1999axisymmetric} led to a series of studies of relativistic pulsar magnetospheres in axisymmetry  \citep{2003ApJ...598..446U,2004MNRAS.349..213G,Gruzinov:2005,Timokhin:2006}. This was followed by time-dependent simulations of dipolar axisymmetric magnetospheres of neutron stars carried out within the framework of relativistic magnetohydrodynamics \citep{komissarov2006simulations}, three-dimensional models describing the magnetosphere of oblique and orthogonal rotators \citep{spitkovsky2006time, kalapotharakos2009three} and resistive force-free electrodynamics (FFE) \citep{Kalapotharakos:2012}.

A key element of the above models is that field-lines crossing the light-cylinder are swept back, they are thus twisted. This ensures that the transition through the light-cylinder is smooth, thus solving one of the major hurdles of early modelling efforts \citep{1979MNRAS.188..385M,1985MNRAS.217..443M}. Yet, these models do not include any twist along magnetic field-lines that do not cross the light cylinder. Twisted magnetospheres may be relevant to magnetars which are neutron stars with anomalous X-ray emission and Soft Gamma-Ray repeating events \citep{2015RPPh...78k6901T,2017ARA&A..55..261K}. Their strong magnetic field powers these events \citep{duncan1992formation, 1995MNRAS.275..255T,1996ApJ...473..322T,2015RPPh...78k6901T,2017ARA&A..55..261K,2018ASSL..457...57G,esposito2021magnetars}. These events release electromagnetic energy through bursts, outbursts and flares \citep{1979Natur.282..587M,1999Natur.397...41H,2005Natur.434.1098H,woods2004soft,2018MNRAS.474..961C}, and are often also accompanied by a variation of their timing properties \citep{2002ApJ...576..381W,2015ApJ...800...33A,2016MNRAS.458.2088P,2017ApJ...841..126S,2019AN....340..340H,2019MNRAS.488.5251L}.

As both the radiative profile and spin-down rate of neutron stars are determined by the magnetic field structure, variations of the magnetic field will affect them both. Moreover, the amount of energy stored within the magnetosphere which can be rapidly released in powering the aforementioned energetic events is altered. This can be achieved by twisting the magnetic field-lines, through shearing of their foot-points. Models of such twisted and sheared magnetic fields have been presented in the astrophysical literature  \citep{1994A&A...288.1012A,lynden1994self,wolfson1995shear,gourgouliatos2008self} with particular focus on magnetar flares and outbursts \citep{thompson2002electrodynamics,beloborodov2007corona,pavan2009topology,glampedakis2014inside,akgun2016force,kojima2017axisymmetric,Akgun:2017ggw,2019MNRAS.484L.124C}. The basic idea of these models is that the magnetic field can accommodate a maximum amount of twist, beyond which it becomes unstable and releases its energy in an explosive manner, possibly through ideal or resistive instabilities \citep{2003MNRAS.346..540L,2006MNRAS.367.1594L,beloborodov2012mechanism,2019MNRAS.489.3769T,2019MNRAS.490.4858M}. These mechanisms are typically studied for non-rotating magnetospheres. This is justified if most of the activity remains close to the stellar surface where the effects of rotation are minimal. However, as it has been shown in numerical modelling  \citep{parfrey2012twisting,2013ApJ...774...92P}, the relativistic effects cannot be neglected if the twist is strong enough, since the poloidal current may occupy a large fraction of the magnetosphere that extends up to the light-cylinder. 

A magnetar field can be twisted either through shearing of the stellar crust that is driven by internal magnetic stresses in the star \citep{Lander:2019,2021MNRAS.506.3578G}, or through currents that flow from the crust to the magnetosphere \citep{2014MNRAS.445.2777F,2019MNRAS.490.4858M}. The flow of poloidal currents within an axisymmetric magnetosphere, implies the presence of a toroidal field. If a closed magnetic field-line, i.e.~a field-line whose both footpoints are anchored on the crust, is differentially sheared a toroidal field will develop. On the contrary, an open field-line, i.e.~a field-line whose one footpoint is anchored on the crust and extends to infinity,  cannot be twisted permanently from action originating at the star as the twist will relax beyond the light-cylinder, leading only to transient behaviour \citep{2020ApJ...897..173B}. In this paper we impose a toroidal field that arises from the azimuthal shearing of the closed field-lines. Open field-lines contain only the poloidal current that is required for their smooth crossing of the light-cylinder, which cannot be externally imposed.

The plan of the paper is as follows. In section~\ref{Math_Formulation}, we present the mathematical setup that describes a twisted magnetosphere under the force-free condition. 
We present the results of the numerical solution in section~\ref{Results}, and discuss them in section~\ref{DISCUSSION}. Their astrophysical applications are presented in section~\ref{APPLICATIONS}, and we conclude with section ~\ref{CONCLUSION}.

\section{Mathematical formulation}

\label{Math_Formulation}

In this section we provide the mathematical equations that describe axisymmetric ideal force-free magnetospheres. 
Assuming that the electromagnetic forces dominate gravitational and inertial ones, as well as pressure gradients, a system in equilibrium satisfies the condition of zero Lorentz force, namely  
\begin{equation}
    \rho_q \bm{E} + \frac{1}{c}\bm{J}\times\bm{B} = \bm{0}\, ,
    \label{Force-free_Rel}
\end{equation}
where $\bm{E}$ is the electric field, $\bm{B}$ the magnetic field, $\rho_q$ the electric charge density, $\bm{J}$ the electric current density and $c$ the speed of light. 
In what follows we will work in cylindrical coordinates $(R,\phi,z)$, and will consider an axisymmetric configuration. Thus, the magnetic field can be expressed in terms of two scalar functions $\Psi(R,z)$ and $I(R,z)$:
\begin{equation}
  \bm{B} = \nabla \Psi \times \nabla \phi + I \nabla \phi\, ,
  \label{Magnetic_Field}
\end{equation}
where, $\nabla\phi\equiv \hat{\phi}/R$.
The function $\Psi(R,z)$ is the poloidal flux and $I(R,z)$ is proportional to the poloidal electric current contained within a circle of radius $R$ whose centre lies on the axis of symmetry of the system at distance $z$ from the horizontal plane. 

The velocity is given by the expression:
\begin{eqnarray}
\bm{v} = \Omega R ~\hat{\bm{\phi}}\,.
\label{eqn:14}
\end{eqnarray}
where $\Omega$ is the angular frequency. Assuming ideal MHD, we obtain the electric field by Ohm's law:
\begin{eqnarray}
\bm{E} = -\frac{\bf{v}}{c} \times \bm{B}\,.
\label{eqn:15}
\end{eqnarray}
Taking the divergence of the electric field we obtain the electric charge density. Then we substitute into equation (\ref{Force-free_Rel}), and by appropriately normalising lengths so that the light-cylinder radius is at $R=R_{LC} \equiv c/\Omega=1$,  we obtain the following equation: 
\begin{equation}
     (1 - R^2)\left(\frac{\partial^2\Psi}{\partial R^2} - \frac{1}{R}\frac{\partial\Psi}{\partial R} + \frac{\partial^2\Psi}{\partial z^2}\right) -2R\frac{\partial\Psi}{\partial R} = -I(\Psi)\frac{dI(\Psi)}{d\Psi}\,,
     \label{FF_R}
\end{equation}
which is the axisymmetric pulsar equation derived by \cite{scharl_wag}. 

In the standard solution of the pulsar equation \citep{contopoulos1999axisymmetric}, the magnetosphere comprises of two parts: a region where both ends of the field-lines are anchored onto the surface of the star (the region of closed field-lines), where it is postulated that $I(\Psi)= 0$ and a region containing the field-lines that cross the light-cylinder and open up to infinity (the region of open field-lines), where in general $I(\Psi)\neq 0$. The functional form of $I(\Psi)$  in that region is prescribed by the requirement that field-lines cross the light-cylinder smoothly.

The condition that $I(\Psi)=0$ for the closed field-lines is based on the assumption that the field of the pulsar does not have any intrinsic twist. While this assumption corresponds to a magnetic field of minimal complexity, the idea of twisted magnetic fields is commonplace in magnetar models. Solutions with non-zero $I(\Psi)$ in the closed field-lines could be relevant and applicable to strongly magnetised systems. Thus, we study systematically systems with non-zero $I(\Psi)$ within the closed region.

 We integrate numerically equation (\ref{FF_R}) within an orthogonal domain of the northern half-space of the system $R\times z \in [0, R_{max}]\times [0, z_{max}]$ excluding the star, which is assumed to have a radius of $r_{NS}$, thus $R^2+z^2 > r_{NS}^2$. In that domain, we apply the standard boundary conditions of the axisymmetric pulsar equation. 
For $R = 0$, that is the $z$-axis, the following relation holds:
\begin{equation}
    \Psi(R = 0, z) = 0
    \label{eqn:17}
\end{equation}
while on the surface of the star $r_{NS}=\sqrt{R^2+z^2}$, the magnetic flux function is that of a dipole:
\begin{equation}
    \Psi(r_{NS}) = \frac{R^2}{\left(R^2 + z^2\right)^{3/2}}\,.
    \label{eqn:18}
\end{equation}
In the numerical setup, we place the inner boundary at $r_{NS}= 0.1R_{LC}$, which, in physical units, implies a rotation angular velocity $\Omega=3000 \mathrm{rad~s^{-1}}$ and a spin frequency of $500 \mathrm{Hz}$. 
Because of North-South symmetry, the vertical derivative of the magnetic flux at the equator for the closed field-lines is equal to zero, namely
\begin{equation}
    \left.\frac{\partial\Psi}{\partial z}\right|_{R<R_{Y},z = 0} = 0\, .
    \label{eqn:19}
\end{equation}
Here, $R_{Y}$ is the radial extent of the region of closed field-lines, namely the cylindrical radius where the last closed field-line crosses the equator. That position is often called the Y-point.
Beyond the last closed field-line, the system forms an equatorial current sheet, thus the boundary condition there is given by the following expression:
\begin{eqnarray}
\Psi(R\geq R_{Y}, z=0)= \Psi(R_{Y}, z=0)\,.
\label{eqn:20}
\end{eqnarray}
In most studies, $R_{Y}$ is typically the light-cylinder radius, yet it has been recently argued that it is likely to be located at around $0.9$ of the light-cylinder for the untwisted magnetosphere \citep{2023arXiv230910482C}. In the twisted case, enforcing the inner edge of the current sheet at the light-cylinder is too restrictive for some of the cases studied and has to be relaxed to allow for physically acceptable solutions.

As the pulsar equation has a critical point at the light-cylinder, we demand that the magnetic field is non-singular and continuous there, thus the magnetic field-lines cross the light-cylinder smoothly. This leads to Dirichlet and Neumann boundary conditions:
\begin{align}
    \Psi(R = 1^+ , z) & = \Psi(R = 1^- , z)
    \label{eqn:21} \\\
    \left.\frac{\partial\Psi}{\partial R}\right|_{R=1^-} & = \left.\frac{\partial\Psi}{\partial R}\right|_{R=1^+} = \frac{1}{2}II'|_{\Psi(R = 1 , z)} \,,
    \label{eqn:22}
\end{align}
where a prime denotes differentiation with respect to $\Psi$. The above relations allow us to determine the functional form of $I(\Psi)$ for $\Psi<\Psi_0$, where $\Psi_0$ is the flux function corresponding to the last open magnetic field-line. In the region of the closed magnetic field-lines ($\Psi>\Psi_0$) the electric current function is set to the following form:
\begin{equation}
    I(\Psi) = \alpha (\Psi(R,z) -\Psi_0)^n\,,
    \label{eqn:28}
\end{equation}
where $\alpha, n$ are parameters that determine the details of the electric current distribution. This permits the inclusion of twist in the closed magnetic field-lines. 

In principle, one could have chosen a functional form for $I(\Psi)$ where the toroidal field does not vanish at the first closed field line located at $\Psi=\Psi_0$. The reason we opted for $I(\Psi_0)=0$ is the following:
In the standard untwisted pulsar magnetosphere, a  discontinuity on the toroidal field appears across the separatrix between the open and closed field-lines. This discontinuity is accompanied by a poloidal current sheet, through which most of the current that flows through the polar cap and crosses the light-cylinder returns to the star.
Therefore, most of the poloidal electric current that is contained at the open field-line region closes along the separatrix between the open and closed field-lines. 
As the toroidal magnetic field is equal to zero there, the modulus of the discontinuity of the poloidal current across the separatrix is equal in the north and the south hemisphere and the system maintains a symmetry across the equator. 

Once we include a toroidal field in the closed-field line region, this could, in principle, interact asymmetrically with the separatrix.  More precisely, while the toroidal field of the open field-lines just outside the separatrix has opposite directions in the northern and southern hemisphere, the toroidal field that we impose in the closed field-lines has the same polarity. This north-south asymmetry of the toroidal field in the closed-line region would, in general, cause a north-south asymmetry in the poloidal field configuration if the toroidal field just inside the separatrix is non-zero. The discontinuity of the toroidal field  determines the intensity of the poloidal current sheet flowing along the separatrix. Thus, for the hemisphere where the toroidal field has the same direction across the separatrix, the discontinuity will be milder, whereas in the other one it will be steeper. Consequently, the poloidal return current sheet will be different in the northern and southern separatrices. To avoid such complications, we set the toroidal field at the edge of the closed field-line region ($\Psi=\Psi_0$) equal to zero, as it is evident from equation (\ref{eqn:28}). This, ensures that the north-south symmetry of the poloidal magnetic field configuration is not broken. Otherwise, the toroidal field in the closed-line region {\it is not} north-south (anti)symmetric.

At the external boundary of the domain ($z=z_{max}$ and $R=R_{max}$) we implement the split monopole boundary conditions, as follows: 
\begin{align}
    \left.\frac{\partial \Psi(R,z)}{\partial z}\right|_{z=z_{max}}&=-\left.\frac{\partial\Psi(R,z)}{\partial R}\right|_{z=z_{max}^-}\frac{R}{z_{max}}
    \\\
     \left.\frac{\partial \Psi(R,z)}{\partial R}\right|_{R=R_{max}}&=-\left.\frac{\partial\Psi(R,z)}{\partial z}\right|_{R=R_{max}^-}\frac{z}{R_{max}}\,.
\end{align}

Thus, the open magnetic lines become radial. These conditions hold at an infinite distance from the surface of the star. However, we apply them at the outer boundary of the integration domain. We have found that they have minimal impact on our calculation, when we increased the integration domain by a factor of two, the solution changed by less than $1\%$. 

\section{simulations and results}

\label{Results}

\subsection{Magnetic field structure}
 
We have integrated equation (\ref{FF_R}) for various choices of the parameters $n$ and $\alpha$ appearing in equation (\ref{eqn:28}) that determine the electric current flowing along the closed magnetic field-lines. We have used a typical $(R,z)$ resolution of $400 \times 1200$, and we have also doubled the resolution in several cases to ensure the convergence of our calculations. We have solved the above equation applying the simultaneous relaxation method in an appropriately modified version of the numerical code used in \cite{gourgouliatos2019coupled}.  In the open field-line region, the value of $I(\Psi)$ is determined by enforcing the conditions described in equations (\ref{eqn:21}) and (\ref{eqn:22}). At the closed field-lines we input the prescribed function for the electric current, given by equation (\ref{eqn:28}), and we apply the elliptic solver. This process is repeated until the system converges to the solution of the system. The current flowing in the region of the open field-lines should formally return through a current sheet on the equatorial plane and the separatrix between the open and closed field-lines. As it is not possible to handle numerically an infinitesimal return current sheet in a finite difference code, we approximate it using a Gaussian function centred at $\Psi=0.975\Psi_0$ and a width of $0.01 \Psi_0$. This amplitude of the Gaussian is chosen so that the entire current closes through the star. Thus, in this narrow layer we do not apply the conditions of equations (\ref{eqn:21}) and (\ref{eqn:22}) but the above predetermined expression.

\begin{figure*}
     a\includegraphics[width=.49\textwidth]{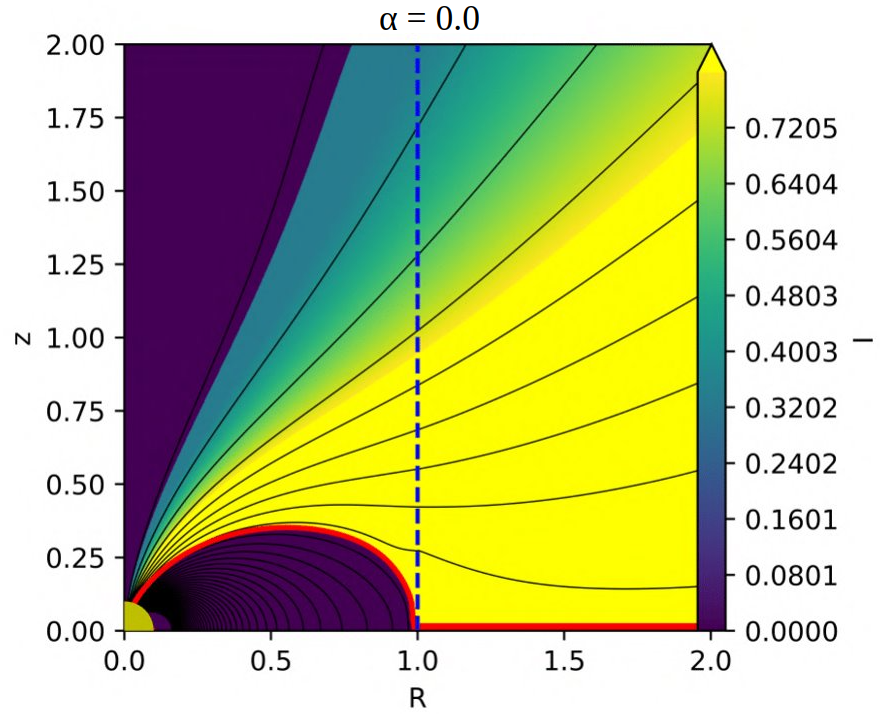}\hfill
     b\includegraphics[width=.48\textwidth]{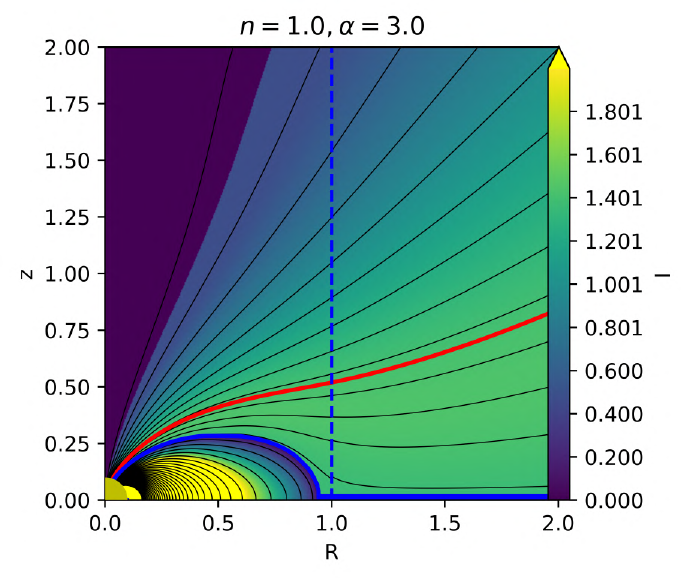}\hfill
    c\includegraphics[width=.48\textwidth]{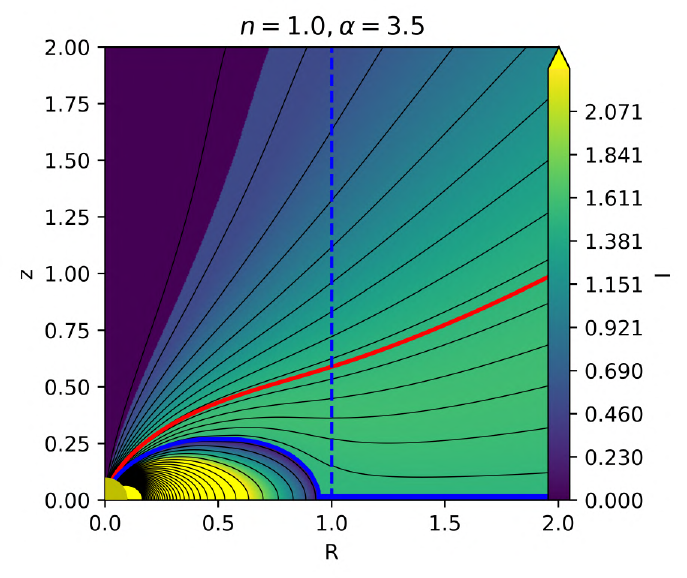}\hfill    
    d\includegraphics[width=.48\textwidth]{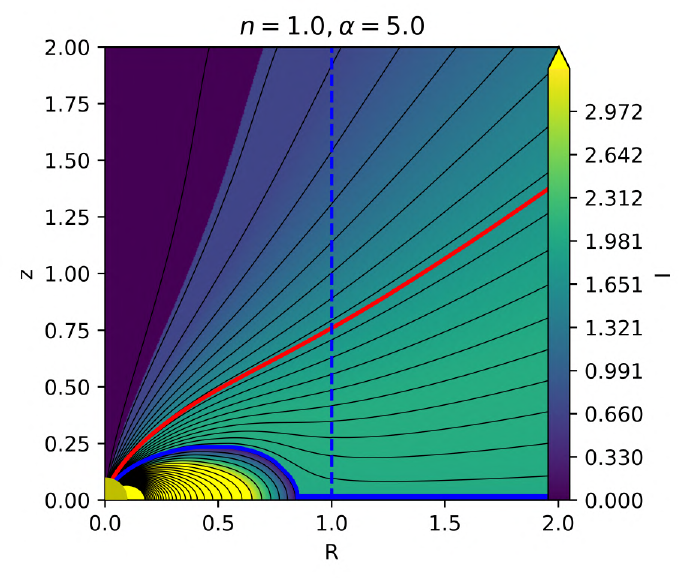}\hfill
    \caption{Pulsar magnetospheres with a linear poloidal current distribution for various values of $\alpha$, ranging from $\alpha = 0$ to $\alpha = 5$. The contours of constant $\Psi$ (poloidal field-lines) are shown in black, and the poloidal current $I$ is shown in colour.  The red line corresponds to the last open field-line of the standard pulsar solution $\Psi_0=1.23$.} 
    \label{fig:g}
\end{figure*}

First, we have tested our numerical code for a system without any poloidal current at the closed magnetic field-line region ($\alpha=0)$ to ensure it converges to the standard pulsar solution, figure \ref{fig:g}, panel a. We find that the last open field-line occurs at $\Psi_0=1.23$ in agreement with the results of \cite{Timokhin:2006}. Next, we have explored a wide variety of combinations of the parameters  $\alpha$ and $n$ shown in Tables \ref{fig:table_1}-\ref{fig:table_3}. 
\begin{table*} 
\caption{ Simulation results for a range of values of the parameter $\alpha$ and $n = 1.0$. The first column is $\alpha$, subsequent columns are $\Psi_0$, the magnetic flux of the last open field-line, $L_{twisted}/L_{untwisted}$ the ratio of the spin-down luminosity of a magnetosphere with twist in the closed field-line region $L_{twisted}$ to a standard `pulsar' solution with twist only in the open field-lines $L_{untwisted}$, $\Delta \phi$ the twist of the closed field-lines and $R_{Y}$ the inner radius of the current sheet. }
\label{fig:table_1}
\begin{tabular}{ |p{0.8cm}|p{1.7cm}|p{1.9cm}|p{1.7cm}|p{2.0cm}| }
 \hline
 \multicolumn{5}{|c|}{n = 1.0} \\
 \hline
 $\alpha$& $\Psi_0$& $L_{twisted}/L_{untwisted}$ & $\Delta\phi$& $R_{Y}$ \\
 \hline
 0.0  & 1.23 & 1.0 & 0.0 & 1.00\\
\hline
0.1  & 1.24  & 1.02 & 0.29 & 1.00 \\
\hline
0.5  & 1.25  & 1.03   &0.31  & 1.00\\
\hline
1.0  & 1.28   & 1.09 &   0.41&        1.00\\
\hline
1.5  & 1.33  &  1.18&   0.47&    1.00\\
\hline
2.0  & 1.41  & 1.32&   0.52&  1.00\\
\hline
2.5  & 1.51   & 1.52  & 0.58  & 1.00\\
\hline
3.0  & 1.64 & 1.78   & 0.64  &  0.95\\
\hline
3.5  & 1.79 & 2.13 & 0.70 &   0.95\\
\hline
4.0  &1.97   & 2.59  & 0.75&   0.90\\
\hline
4.5  & 2.26   & 3.40& 0.81  &  0.90\\
\hline
5.0  & 2.40  & 3.81 & 0.83 & 0.85\\
\hline
8.0  & 3.14  & 6.53 &0.97  & 0.50\\
\hline
10.0  &3.64   & 8.77 & 1.08 & 0.45\\
\hline

\end{tabular}
\end{table*}

The linear current relationship corresponds to $n=1$, and we have solved for $\alpha=0.1$ to $\alpha=10$. In figure \ref{fig:g} we show solutions for the parameters starting from $\alpha = 0$, the untwisted case, to $\alpha = 5.0$, panels a to d. As $\alpha$ increases closed magnetic lines of the untwisted system open up gradually, which is reflected on the increase of $\Psi_0$, that is found through the solution of the pulsar equation. Increasing $\alpha$ above $2.5$, while enforcing the innermost point of the current sheet to remain at $R_{Y}=1$ leads to magnetospheric structures that contain magnetic field-lines that are disconnected from the star. While such solutions satisfy the mathematical form of the equations and the boundary conditions, they are physically unacceptable, as they include field-lines emerging from the equator without being linked to the star. To prevent such unphysical solutions we move the innermost point of the current sheet within the light-cylinder and closer to the surface of the star. This way, the current sheet starts earlier and all magnetic field-lines remain connected to the star. While we set manually the position of $R_Y$, this is not done arbitrarily, but in a way to impose the minimum displacement of $R_Y$ to avoid the appearance of disconnected magnetic field lines. For $\alpha = 3.0$, the current sheet starts at $R_{Y} = 0.95$. 
The next shift of the current sheet occurs for $\alpha$ = 4.0 where  the current sheet inner edge must be shifted to $R_Y = 0.9$. We have used increments of $0.05$ for the displacement of $R_Y$, in $R_{LC}$ units. Should we have allowed for a finer change, the movement of $R_Y$ would be continuous with $\alpha$.
In table \ref{fig:table_1}  we can see the increase in the value of the magnetic flux function at the last open magnetic field-line with respect to the coefficient $\alpha$ and the displacement of the current sheet. Such a behavior is expected, as moving $R_{Y}$ closer to the star leads to larger values of the flux function. This, in combination with the increase of $\alpha$ leads to a nonlinear increase of the magnetic flux function in the corresponding last open magnetic line.

\begin{figure*}
    a\includegraphics[width=.48\textwidth]{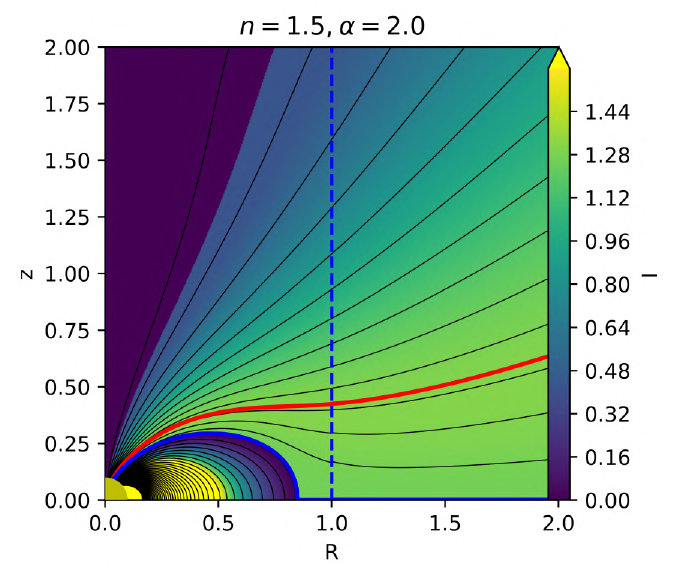}\hfill
    b\includegraphics[width=.48\textwidth]{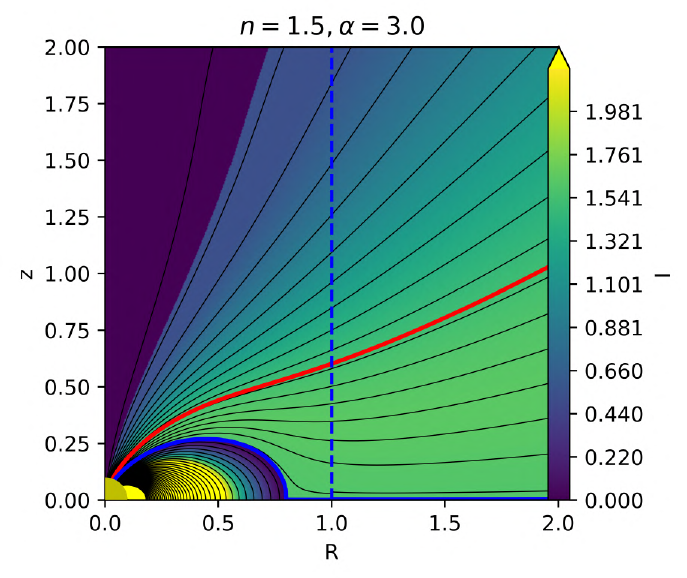}\hfill
    c\includegraphics[width=.48\textwidth]{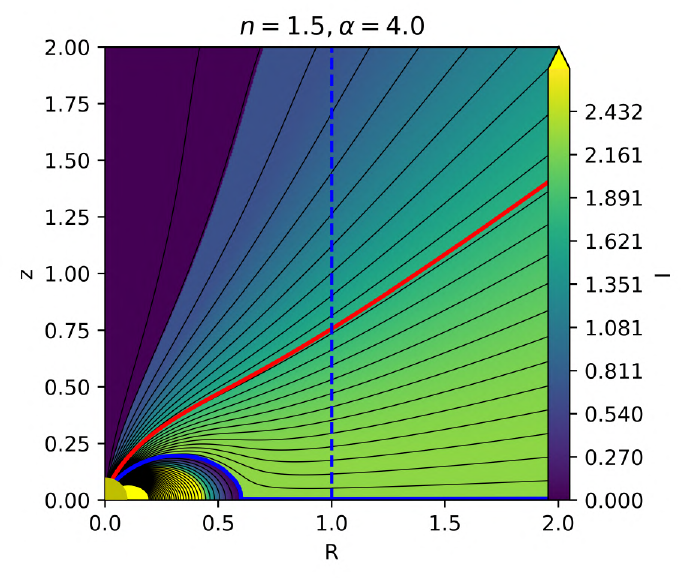}\hfill
    d\includegraphics[width=.48\textwidth]{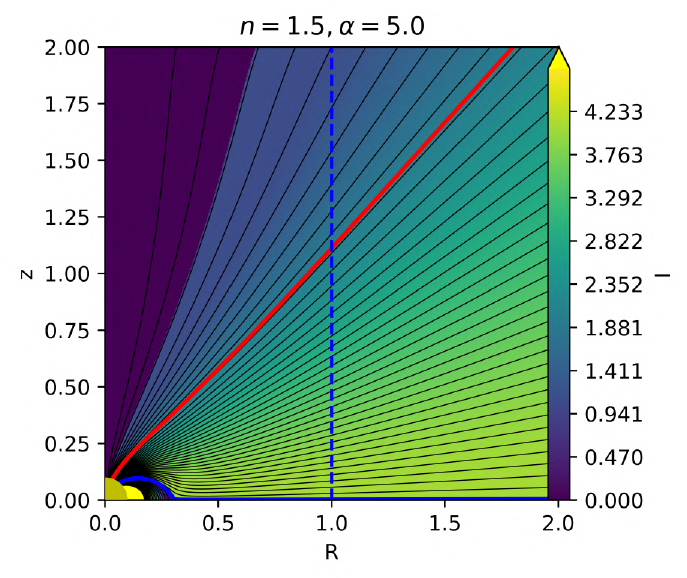}\hfill
    \caption{Magnetospheres including a non linear poloidal current distribution (n = 1.5) for values of $\alpha$ ranging from $\alpha = 2.0$ to $\alpha = 5$. The contours of constant $\Psi$ (poloidal field-lines) are shown in black and in color $I$. The red line corresponds to the last open field-line of the standard pulsar solution $\Psi_0=1.23$.}
    \label{fig:o}
\end{figure*}

The same behaviour is found for the nonlinear dependence of the current  distribution to the magnetic flux function. In particular, we have studied the magnetosphere for $n = 1.5$ and $\alpha$ ranging from 0.1 to 10, with some solutions shown in Figure \ref{fig:o}. We can obtain solutions with the current sheet inner edge being at $R_{Y}=1$, while  $\alpha \leq 1.5$.  Once $\alpha$ reaches $2.0$ 
a drastic shift of the inner edge of the current sheet is required to $R_{Y}=0.85$.
Here also, the red line denotes the magnetic field-line for which $\Psi_0= 1.23$. The increase of the coefficient $\alpha$ is followed by a stronger shift of the current sheet towards the star reaching $R_Y = 0.40$ for $\alpha = 5.0$.
Similar to the linear case, 
the magnetic energy of the region of the closed magnetic lines 
increases initially with $\alpha$.  
The change in the position of the current sheet also leads to a change in the volume of the occupied by the closed magnetic field-lines and eventually a decrease 
of the magnetic energy.

\begin{table*} 
\caption{Simulation results for a range of values of the parameter $\alpha$ and $n = 1.5$.}
\label{fig:table_2}
\begin{tabular}{ |p{0.8cm}||p{1.7cm}|p{1.9cm}|p{1.7cm}|p{2.0cm}|  }
 \hline
 \multicolumn{5}{|c|}{n = 1.5} \\
 \hline
 $\alpha$& $\Psi_0$& $L_{twisted}/L_{untwisted}$& $\Delta\phi$& $R_{Y}$ \\
 \hline
0.1  & 1.24 & 1.02 & 0.29 & 1.00\\
\hline
0.5  & 1.26 &1.05   &0.36  &  1.00\\
\hline
1.0  & 1.30  &  1.13&   0.51&        1.00 \\
\hline
1.5  & 1.40 &  1.30&   0.57&        1.00  \\
\hline
2.0  & 1.59  & 1.68&   0.60&  0.85         \\
\hline
2.5  & 1.83   & 2.23  & 0.69  &       0.85    \\
\hline
3.0  & 2.20  & 3.2    & 0.99  &  0.80 \\
\hline
3.5  & 2.56   &4.32   & 1.11  &  0.70   \\
\hline
4.0  & 2.91  & 5.62    & 1.20 & 0.60            \\
\hline
4.5  & 3.25 &6.99& 1.31   & 0.50    \\
\hline
5.0  &3.65 & 8.81 & 1.38  & 0.40   \\
\hline
8.0  &4.24 & 11.8 &  1.44 & 0.30   \\
\hline
10.0  & 4.62& 14.1& 1.50 & 0.30 \\
\hline
\end{tabular}
\end{table*}

Qualitatively similar results can be drawn studying the non-linear current distribution for $n=2.0$ and $\alpha$ ranging from $0.1$ to $10$. In this case, the inner edge of the current sheet has to be displaced even more drastically compared to the other two cases. 
\begin{figure*}
    a\includegraphics[width=.48\textwidth]{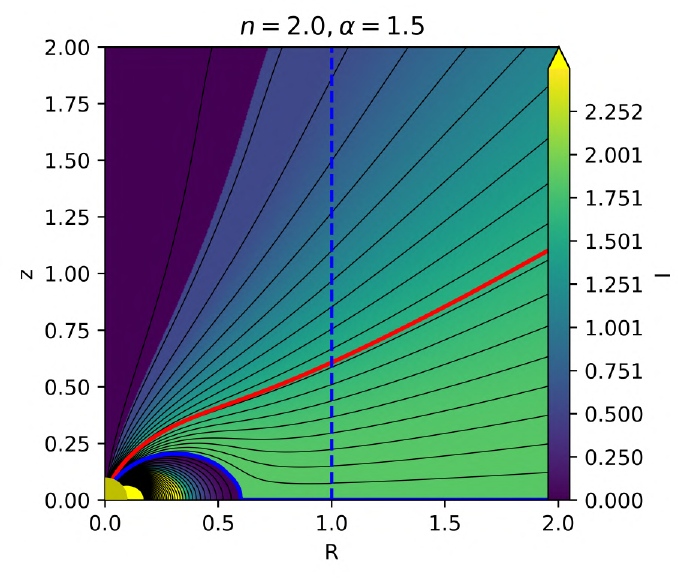}\hfill
    b\includegraphics[width=.48\textwidth]{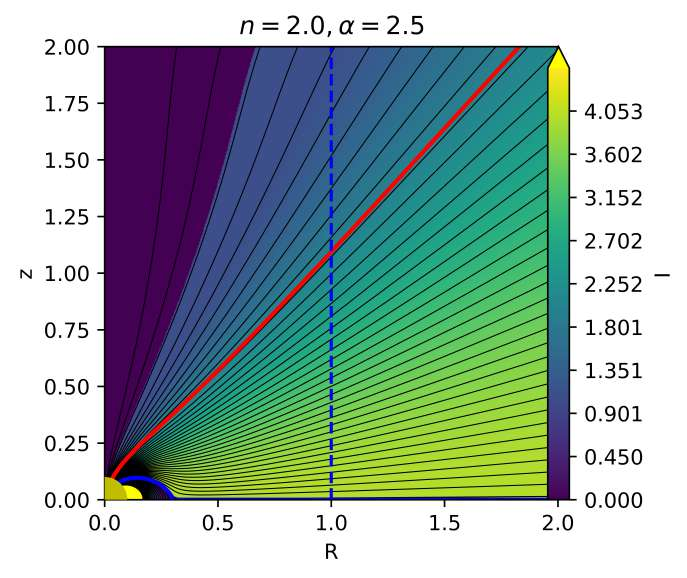}\hfill
    c\includegraphics[width=.48\textwidth]{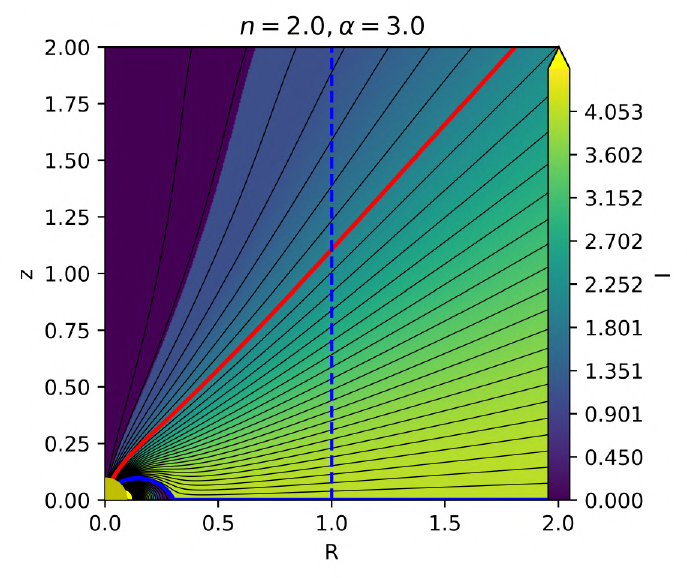}\hfill
    d\includegraphics[width=.48\textwidth]{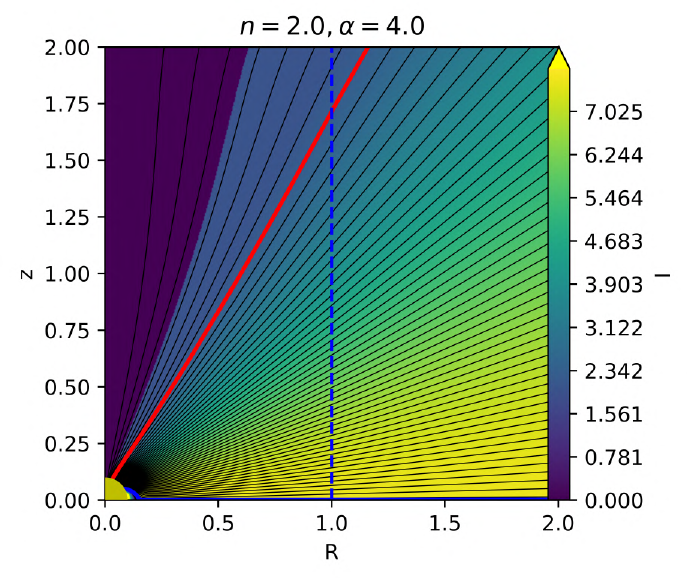}\hfill
    \caption{Magnetospheres under the influence of a non linear polar current distribution (n = 2.0) for values of $\alpha$ ranging from $\alpha = 1.5$ to $\alpha = 4.0$. The contours of constant $\Psi$ (poloidal field-lines) are shown in black and in color $I$. The red line corresponds to the last open field-line of the standard pulsar solution $\Psi_0=1.23$. }
    \label{fig:p}
\end{figure*}
\begin{table*}
\caption{ Simulation results for a range of values of the parameter $\alpha$ and $n = 2.0$.}
 \label{fig:table_3}
\begin{tabular}{ |p{0.8cm}||p{1.7cm}|p{1.9cm}|p{1.7cm}|p{2.0cm}|  }
 \hline
 \multicolumn{5}{|c|}{n = 2.0} \\
 \hline
 $\alpha$& $\Psi_0$& $L_{twisted}/L_{untwisted}$& $\Delta\phi$& $R_{Y}$ \\
 \hline
0.1  & 1.24  & 1.03 & 0.3 & 1.00\\
\hline
0.5  & 1.29   & 1.09   &0.39  & 1.00\\
\hline
1.0  & 1.53 &  1.56    &   0.60&   1.00\\
\hline
1.5  & 2.31  & 3.52&   0.67&  0.60\\
\hline
2.0  &3.29   & 7.19&   0.70&  0.40\\
\hline
2.5  & 3.80    & 9.58    & 0.89  &    0.35\\
\hline
3.0  & 4.40   & 12.77    & 1.12  &    0.30\\
\hline
3.5  & 5.03  & 16.74   & 1.13  &        0.25\\
\hline
4.0  & 5.12   & 17.58    & 1.25 &         0.20\\
\hline
4.5  & 6.11  & 24.71& 1.39   &   0.15\\
\hline
5.0  & 6.16& 25.13 & 1.65  &  0.15\\
\hline
8.0  &7.21 &34.36  & 1.70&  0.15\\
\hline
10.0  &8.76 &50.72  &  1.75& 0.13\\
\hline
\end{tabular}
\end{table*}
Figure \ref{fig:p} shows the magnetic field structure for a poloidal current distribution with $n = 2.0$ for various values of $\alpha$. The value of $\alpha = 1.0$ corresponds to the maximum achievable value of the current function with $n = 2.0$, for which no displacement of the current sheet from the position $R_{Y}=1$ is required. For the highest value $\alpha=10$ we have studied, we found that $R_Y=0.13$ and the system is very close to the structure of a split monopole. 

\begin{figure}
    \centering
    \includegraphics[width=0.50\textwidth]{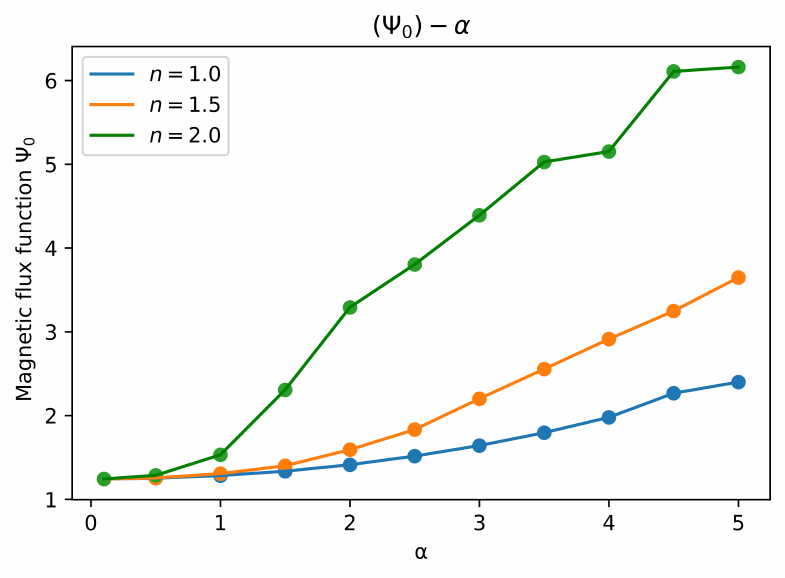}
\caption{Magnetic flux function of the last closed magnetic line for a variety of current distributions. }
 \label{fig:r}
\end{figure}

\subsection{Spin-down luminosity and twist of the magnetosphere}

Introducing a poloidal current in the closed field-line region, and increasing its intensity, results 
in more magnetic field-lines becoming open, Figure \ref{fig:r}. 
As the spin-down rate is related to the torque exerted by the open magnetic field-lines, such an increase 
leads to a more efficient loss of the rotational energy of the star. At a given rotation rate, the spin-down luminosity $L$ scales approximately as follows:
\begin{equation}
    L \approx\frac{4}{3}\Psi_{0}^2\sim\Psi_{0}^2.
    \label{eqn:29}
\end{equation}
We note that a more accurate approximation of energy losses can be obtained from the relation \citep{Timokhin:2006}:
\begin{equation}
    L = 2\int_0^{\Psi_0}  I(\Psi)d\Psi
    \label{eqn:100}
\end{equation}
(the total spin-down luminosity is equal to the electromagnetic energy loss from both poles of the star). We have found that the difference of the results found by the two formulae are minimal and in the results we quote the ones that scale as $\Psi_0^2$.
From this we are able to obtain an estimate of the energy loss of the star under the influence of the increasing current in the region of the closed magnetic lines. Increasing the value of the current in the region of the closed magnetic lines is followed by an increase spin-down rate which is more effective for higher $n$. It is interesting to see how the star loses part of its rotational energy in the case where the region of closed magnetic lines in its magnetosphere is driven by a non-linear current distribution, with n = 2.0. In this case, we can observe a significantly steeper increase in the energy loss as the value of $\alpha$ increases compared to the previous two cases. It is also possible to discern sudden increases in the spin down power which also  coincide with the respective displacement of the current sheet towards the inner magnetosphere of the star. If this displacement is significant, equation \eqref{eqn:29} may substantially underestimate the true spin-down rate; see section \ref{twist_spindown}. 

The equation of a field-line can be found by integrating the following relation: 
\begin{equation}
    \frac{dR}{B_R} = \frac{R d\phi}{B_{\phi}} = \frac{dz}{B_z}
    \label{eqn:30}
\end{equation}
\begin{figure}
    \centering
    \includegraphics[width=0.48\textwidth]{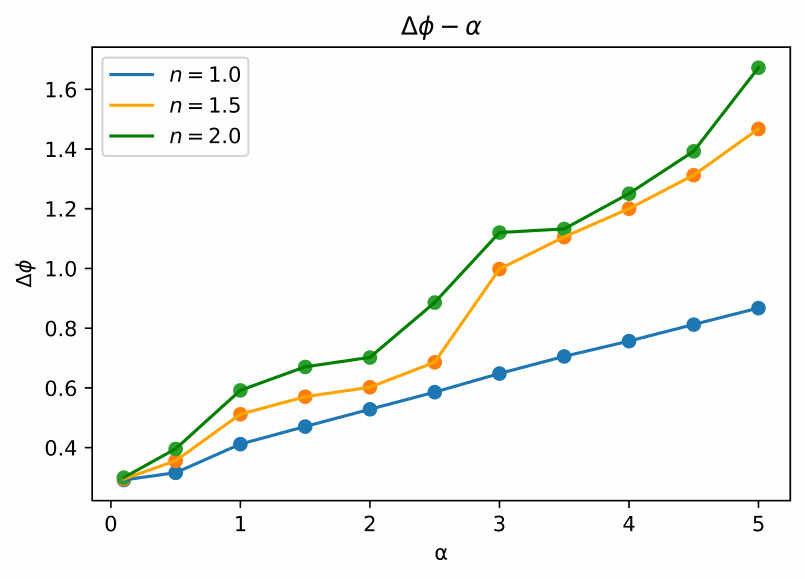}
\caption{An approximation of how the magnetosphere in the region of the closed magnetic field-line twists, as the value of the current function increases in the closed magnetic
lines of its magnetosphere and as the current sheet shifts towards its surface.}
\label{fig:v}
\end{figure}
\begin{figure}
    \centering
    \includegraphics[width=0.50\textwidth]{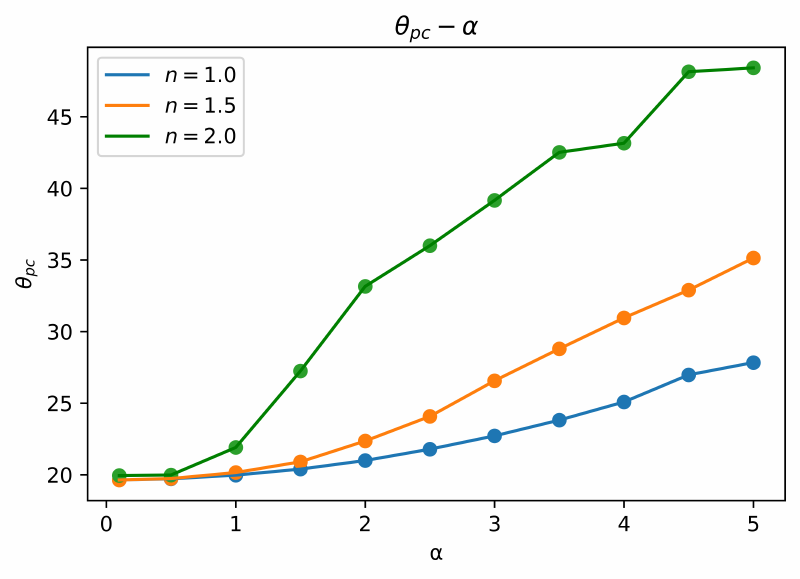}
\caption{An approximation of how the polar cap increases , as the value of the current function increases in the closed magnetic
lines of its magnetosphere and as the current sheet shifts towards its surface.}
 \label{fig:w}
\end{figure}
where $B_R, B_z, B_{\phi}$ are the components of the magnetic field in cylindrical coordinates. We can evaluate the twist of a field-line, through integration of the equation involving the $R$ and the $\phi$ components of the magnetic field. To estimate the maximum possible twist that the closed magnetic field-lines of the magnetosphere can undergo, the calculations were performed within the region of the closed magnetic field-lines at a position corresponding to a magnetic line having a magnetic flux $\Psi = 1.1\Psi_0$. We performed the integration on various field-lines within the closed region and we have found for the particular choices of magnetic configuration the above expression provides the maximum twist. Using the first equality of equation (\ref{eqn:30}) and integrating along a constant $\Psi$ we obtain the following relation: 
\begin{equation}
    \Delta\phi = \int \frac{B_{\phi}}{RB_R}dR .
    \label{eqn:31}
\end{equation}
We integrate from  
a point on the surface of the star in the northern hemisphere, where the field line emanates, to the equator, thus the twists quoted in the tables correspond to half of the field-line's twist.

We plot the results in Figure \ref{fig:v}. 
We notice that for the system where the current sheet is closest to the integration domain inner edge and $\alpha$ has the highest value quoted, ($n=1$ $\alpha =10$; $n=1.5$, $\alpha =10$ and $n=2$, $\alpha=10$), bf the twist does not increase indefinitely but rather tends to approach a maximum value $\sim \pi/2$ which is consistent with previous solutions \citep{lynden1994self,gourgouliatos2008fields,pavan2009topology,Akgun:2017ggw}. Thus, after a finite amount of twist, the system can formally reach a state of a split monopole. In our calculation we have chosen to explore the characteristics in terms of the parameter $\alpha$ going up to a maximum value of $\alpha=10$. Thus while two simulations have the same value of $\alpha$, different $n$ corresponds to different amounts of twist, with the increase in the twist being more drastic for larger $n$ that also require pushing the $Y$ point closer to the star. 

To calculate the opening angle of the polar cap, the boundary condition describing the magnetic flux function on the surface of the star was used, equation (\ref{eqn:18}), which can be written using the polar angle as follows:
\begin{equation}
\Psi_0 =  \frac{\sin^2\theta_{pc}}{R_{NS}} 
\label{eqn:32}
\end{equation}
and consequently, the angle of the polar cap will be given by the relation:
\begin{equation}
    \theta_{pc} = \arcsin\left(\sqrt{\Psi_0 R_{NS}}\right)\,.
    \label{eqn:33}
\end{equation}

In Figure \ref{fig:w} we plot the polar cap angle $\theta_{pc}$ versus $\alpha$. We verify that by increasing the twist the polar cap expands. A further conclusion we can draw is that the position of the current sheet plays a crucial role in shaping the polar cap, which is consistent with its impact on the magnetosphere in general. This claim can become more evident by studying the polar cap 
in the case where the magnetosphere is described by the non-linear current distribution $\alpha(\Psi - \Psi_0)^{2.0}$. In this case, in addition to the very large values obtained for the current function, compared to the other two cases, the current sheet is subject to the strongest currents compared to the other cases. In these magnetospheres we can also observe the largest increases of the polar cap opening angle.

\section{Discussion}
\label{DISCUSSION}

The inclusion of a twisted region in the relativistic pulsar magnetosphere has significant qualitative implications for its structure and main characteristics. We have identified the following changes: displacement of the inner edge of the equatorial current sheet; enhancement of the spin-down rate related to the twist and an increase to the size of the polar cap.

\subsection{Location of the current sheet}

An essential part of the axisymmetric pulsar magnetosphere is the equatorial current sheet. The field-lines that cross the light-cylinder cannot cross the equatorial plane, as this would imply that the particles that are frozen onto them should move faster than the speed of light. Therefore, these field-lines form two discontinuous streams separated by the equator. While the above requirement enforces the presence of a current sheet beyond the light-cylinder, there is no physical reason to prevent the inner edge of the current sheet to be closer to the star and within the light-cylinder. Such solutions are mathematically and physically viable. In most pulsar studies, the location of the current sheet inner edge is set to the light-cylinder, where a $Y-$point appears. Despite that, it has been shown that a family of smooth solutions can be obtained for current sheets starting within the light-cylinder \citep{Timokhin:2006}. In these solutions, there is no physical requirement to demand that the current sheet starts within the light-cylinder. However, as a twist is introduced in the magnetic field-lines, the displacement of the current sheet within the light-cylinder is no longer just an alternative solution, but a necessity in order to obtain a physically acceptable configuration. This is because in the untwisted pulsar magnetosphere, there is only one characteristic length-scale, the light-cylinder. In the twisted magnetosphere however, an additional length scale is introduced, which is inversely proportional to $\alpha$, that determines the ratio of the magnetic field to the electric current flowing along a given magnetic field-line. This becomes evident from the solution of an axisymmetric force-free field in spherical geometry, the so-called spheromak solution \citep{1957ApJ...126..457C}, that has a characteristic radius at which the field-lines close. We further note that such constraints appear in relativistic generalisations of force-free solutions \citep{2010GApFD.104..431G, 2022ApJ...934..140B}. Thus, in the solutions studied in this paper, we see the interplay between these two length-scales: the light-cylinder and the force-free twisted field. As long as the twist is small, and consequently $\alpha^{-1}$ large, the force-free length scale is sufficiently long, the location where the twisted field-lines must formally close lies beyond the light-cylinder. Therefore, the inner edge of the current sheet is not affected. As $\alpha$ increases there will be a point where the length scale of the twisted field will become comparable to the light-cylinder. Once this stage is reached, we obtain solutions where the current sheet inner edge is located inside the light-cylinder. We further note that a similar effect has been noted in studies where additional length scales are introduced, such as the the Alfv\`en radius \citep{10.1111/j.1365-2966.2011.18280.x,2020MNRAS.494.4838L}. Similarly, dynamical studies of magnetospheres lead to transient configurations where the current sheet starts well within the light-cylinder  \citep{2022ApJ...934..140B,2023arXiv230208848S}. Moreover, in solutions obtained through force-free electrodynamics \citep{Spitkovsky:2006}, the magnetosphere very quickly opens up at $R=0.6 R_{LC}$ and it took it more than 20 rotations to gradually reach the light cylinder.

\subsection{Polar cap}

The current density throughout the magnetosphere of a pulsar is fuelled by the creation of positron-electron pairs, possibly in the polar cap region, filling the magnetosphere with charged particles \citep{1975ApJ...196...51R,1978ApJ...222..297S,10.1093/mnras/255.1.61}. As the region of the polar cap has been associated with radiation mechanisms and particle acceleration its physical size may be relevant to pulse shape.  Enhancing the twist of the magnetosphere, leads to an expansion of the polar cap region. This is related to the fact that more field-lines become open. Since the opening angle of the polar cap depends on the region where the last open field-line crosses the star, having a larger fraction of open field-lines implies a larger polar cap. We notice significant increases of the polar cap once it is required to displace the current sheet closer to the surface of the star in magnetospheres with a current sheet near the surface of the star.
\begin{figure}
    \centering
    \includegraphics[width=0.50\textwidth]{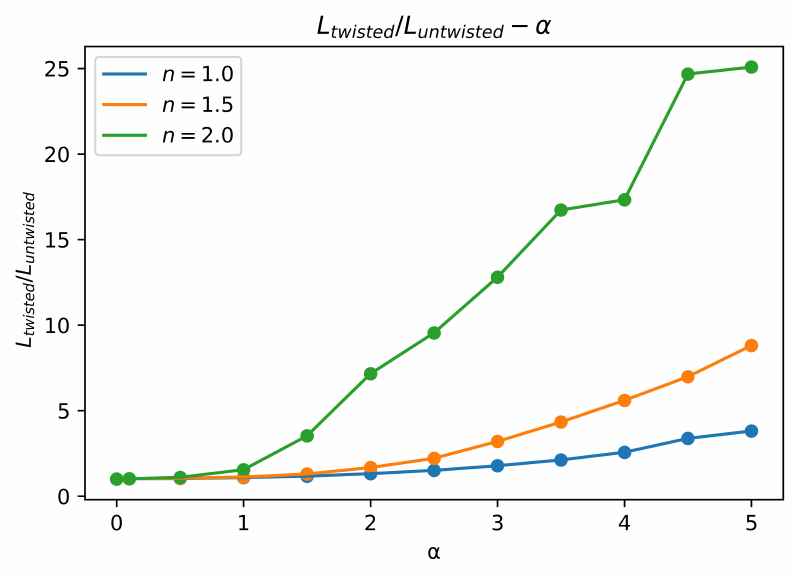}
\caption{The spin-down power of a twisted magnetosphere scaled to the spin-down power of an untwisted one against $\alpha$ for the various values of $n$.}
 \label{fig:z}
\end{figure}

\subsection{Twist and spin-down rate}
\label{twist_spindown}

In general, twisted magnetospheres are expected to have more efficient spin-down rates due to the fact that field-lines tend to become radial and approach the structure of a split monopole \citep{1982RvMP...54....1M,2002ApJ...574..332T,2006MNRAS.367.1594L}. 
An additional factor contributing to more efficient spin-down of twisted magnetospheres is the displacement of the current sheet closer to the star. In figure \ref{fig:z} we plot the spin-down rate as a function of the displacement of the current sheet. It is evident that it increases as the current sheet moves inwards in the magnetosphere. We note however, that while we may have the same displacement of the $Y$-point, the spin-down rate does not become identical in these cases, as the details depend on the functional form of $I=I(\Psi)$ in the closed field-line region, leading to some deviations.   

In fact, our results may underestimate the spindown rate for these cases, since we assume the luminosity scales with $\Psi_0^2$ only. In practice, there may be an additional contribution from the qualitative change to the magnetospheric structure: because both the polar cap region is larger and $R_{Y}$ closer to the star, there is a larger volume of the exterior where electromagnetic waves carry away angular momentum from the star.
It is interesting to compare our results with the case of adding a neutrino-driven wind, which also results in a larger volume of open field-lines. In the semi-analytical analysis of \citet{2020MNRAS.494.4838L} (specifically, combining their equations 2,6, and 12), it may be shown that the spindown luminosity has an additional scaling with $(R_L/R_{Y})^2$ due to the changes in magnetospheric structure. A similar contribution is likely to exist in our case, and would result in lines of steeper gradient in fig. \ref{fig:z}.

\section{Applications}
\label{APPLICATIONS}

While the discussion has been primarily numerical and focusing on the generic features of twisted magnetospheres we can still consider some applications to specific astrophysical systems. The effects described in this work may be related to the switching of neutron stars between pulsar and magnetar states, as this has been observed in PSR J1119-6127 \citep{2017ApJ...834L...2M}. This source has been exhibiting the following behaviour: Two glitches occurred in 2004 and 2007. The glitch events were followed by intermittent and RRAT-type emission; the pulsar recovered from the glitch in an interval between 20 and 210 days. The spin-down rate recovered to a value smaller than the pre-glitch one \citep{2011MNRAS.411.1917W}. Another event occurred in 2016 with a glitch event accompanied by radiative changes \citep{2018MNRAS.480.3584D}. While our model is a series of static axisymmetric equilibrium states and cannot capture this type of event in its totality, some of the characteristics can be described in this framework. A glitch can provide a sudden twist in the magnetosphere, part of which may be channeled to the open field-lines, but a fraction of the twist may be trapped in the closed field-lines \citep{2020ApJ...897..173B}. Following that, a twisted magnetosphere will lead to a higher spin-down rate compared to the pre-glitch state. Should the magnetosphere have some resistivity, the spin-down rate will gradually drop, with a minimum value corresponding to the completely untwisted one. Furthermore, the process of twisting has three effects: an increase in the current flowing within the closed field-line region, the expansion of the polar cap and the displacement of the current sheet within the light-cylinder. The first one may increase the amount of plasma in the closed field-lines and shadow radio pulses, which is actually the case, while particles accelerated in this region may bombard the crust and lead to X-ray emission \citep{Beloborodov:2009}. The displacement of the current sheet and the expansion of the polar cap affects the interface between the closed and open field-lines, which is related to the gaps where the creation of pairs is likely.  Thus, it is likely that the presence of a twisted region of the field-lines can lead to magnetar behaviour to an otherwise normal pulsar, even if it is not highly-magnetised.

We further remark that some pulsars exhibit moding and nulling behaviour, and changes in the subpulses, \citep{1986ApJ...300..540D,2005MNRAS.356...59E,2010MNRAS.408..407B,2013MNRAS.433..445R}. These systems include the intermittent pulsars: B1931+24 \citep{2008MNRAS.391..663R}, J1832+0029 \citep{2012ApJ...758..141L}, J1107-5907 \citep{2018ApJ...869..134M}, J1832+0029 and J1841-0500 \citep{2020ApJ...897....8W}. 
 These sources switch between states with different pulse shape, or they even completely turn off, while at the same time their spin-down frequency changes. These effects have been explained in the context of changes in the size of the corotating region of the magnetosphere and transitions between metastable configurations \citep{2010MNRAS.408L..41T}. While we can construct solutions with different spin-downs, we cannot at this point offer a clear mechanism that would show how the system can switch between these two configurations. However, qualitatively, we may expect the generation of such states: in a  twisted magnetosphere the location of the current sheet is displaced, strongly affecting the closed field-line area and consequently the spin-down rate. Therefore, the system may transition  between states with different spin-down rates. In particular, the intermittent pulsar B1931+24 alternates between an on and an off state, when the radio emission switches on and off respectively, with the on state lasting about 5-10 days, and the off state 25-35. The spin down rate alternates between: $\dot{v}_{on}= (-16.30\pm 0.04)\times 10^{-15} \mathrm{s}^{-2}$ and $ \dot{v}_{off} = (-10.80\pm 0.02)\times10^{-15} \mathrm{s}^{-2}$. Considering the runs with $n=1$, we notice that the states with $\alpha=2$ and $\alpha =3.5$ have spin-down powers whose ratio is $2.13/1.32=1.61$. This ratio is close to the ratio of the spin-down powers of the two states, with the highly twisted one being the on phase and the less twisted one being the off phase.

\section{Conclusions}

\label{CONCLUSION}
The main focus of this paper is the structure of a twisted axisymmetric magnetosphere of a relativistically rotating neutron star, by numerically solving the relativistic force-free equation. 
The increase of the poloidal current in the closed magnetic lines, or equivalently the introduction of a twist in the closed field-lines, affects the global structure of the magnetosphere and it is found to be qualitatively different from the standard (untwisted) pulsar magnetosphere.  
We find that part of the magnetic field-lines  
becomes open. 
Moreover, we find that it is not possible to have a simultaneous increase of the polar current and a physically consistent structure of the magnetosphere without a displacement of the current sheet towards the surface of the star. We conclude that the shift of the current sheet is unavoidable if the current in the magnetosphere increases above a certain value.
During the movement of the current sheet towards the surface of the star, the volume of closed field-lines decreases. Thus, following an initial increase in the energy of the closed field-line, it eventually drops. 
Furthermore, the maximum value of the twist is approximately $\Delta\phi_{max} = \pi/2$, which is in agreement with previous studies, thus after finite twist, the magnetosphere will open up completely adopting the structure of a split monopole. 

It is possible that variations in the timing behavior of magnetars before and after outbursts can be associated with the magnetospheric twist. We conclude that by twisting the closed magnetic field-lines the field becomes more radial, a larger fraction of magnetic flux crosses the light-cylinder and the spin-down rate becomes higher. As the twist cannot be indefinitely increased we anticipate that after some twisting episode the field will return to the untwisted state either through a sudden event, i.e.~reconnection of the open field-lines, or through gradual dissipation of the current.

We note that these results can be further explored in a 3-D study since an axisymmetric system cannot simulate a pulsating source. Even in axial symmetry the inclusion of quadrupolar, or in general multipolar fields, breaks the north-south symmetry. The lack of a north-south symmetry leads to a current sheet that does not coincide with the equator, and would require a drastically different treatment from the finite difference code we are employing here.

While these results are drawn from an idealised case of an axisymmetric rotator, they can provide useful conclusions related to the behaviour of magnetars and neutron stars. For instance, changes in the twist affects the shape of the closed region, the spin down rate and the polar cap. More interestingly, we notice that a pulsar can switch between twisted and untwisted states, without drastically changing the energy content of the closed region. This could be possibly related to intermittent sources, and sources switching from pulsar to magnetar behaviour and vice versa.

\section*{ACKNOWLEDGMENTS}
KNG acknowledges funding from grant FK 81641 "Theoretical and Computational Astrophysics", ELKE.

\section*{Data availability statement}
The data underlying this article will be shared on reasonable request to the corresponding author.

\bibliographystyle{mnras}
\bibliography{main.bib}

\label{lastpage}

\end{document}